\documentclass[letter]{aa} 
\usepackage{graphicx}
\usepackage{txfonts}
\usepackage{natbib}
\usepackage{hyperref}

\def\deg{$^{\circ}$}

\begin{document}

\title{
  VLTI/AMBER unveils a possible dusty pinwheel nebula in
  WR118.\thanks{The observations presented in this paper were
    obtained with the AMBER instrument of ESO's Very Large Telescope
    Interferometer (VLTI) as part of the Guaranteed Time Programme
    079.D-0359(A) (PI: T.~Driebe).}
}
\subtitle{}

\titlerunning{VLTI/AMBER observation of \object{WR\,118}}
\author{F.~Millour\inst{1}
  \and
  T.~Driebe\inst{1}
  \and
  O.~Chesneau\inst{2}
  \and
  J.~H.~Groh\inst{1}
  \and
  K.-H.~Hofmann\inst{1}
  \and\\
  K..~Murakawa\inst{1}
  \and
  K.~Ohnaka\inst{1}
  \and
  D.~Schertl\inst{1}
  \and
  G.~Weigelt\inst{1}
}

\institute{Max-Planck-Institut f\"ur Radioastronomie, Auf dem H\"ugel 69, D-53121 Bonn, Germany, \\
  \email{fmillour@mpifr-bonn.mpg.de}
  \and
  Observatoire de la C\^ote d'Azur/CNRS, UMR 6525 H. Fizeau, Univ. Nice Sophia Antipolis, 
  Avenue Copernic, 06130 Grasse, France
}
\date{Version:\today}


\abstract
{
  Most Wolf-Rayet stars (WR) of WC9 sub-type exhibit a
  dusty circumstellar envelope, but it is still a matter of debate how
  dust can form in their harsh environment. In a few cases, a
  pinwheel-like structure of the dusty envelope has been
  detected. Therefore, it has been suggested that dust formation in
  all dusty WR stars might be linked to colliding winds in a
  binary system.
} 
%
{
  We probed the innermost region of the circumstellar dust shell of the deeply embedded WR star \object{WR\,118}. 
}
{
  We carried out spectro-interferometric observations using the AMBER
  instrument of ESO's Very Large Telescope Interferometer in
  low-spectral resolution mode ($R=35$). The $K$-band observations were
  obtained with three 1.8~m telescopes spanning projected baselines
  between 9.2 and 40.1~m. 
}
{
  At high spatial frequencies, the AMBER visibilities exhibit a
  prominent lobe, indicating that the envelope contains one or several
  zones with a large local intensity gradient. The strong closure phase
  signal clearly shows that the circumstellar envelope of
  \object{WR\,118} can only be described by an asymmetric intensity
  distribution. We show that a pinwheel nebula seen at low
  inclination is consistent with the AMBER data. Its size was
  determined to be $13.9\pm1.1$\,mas.
}
{
  \object{WR\,118} possibly harbors a pinwheel nebula, which
  suggests a binary nature of the system. According to our best
  model, the period of the system would be $\approx60$ days
  (for $d=3$\,kpc), making WR\,118 the shortest-period pinwheel nebula
  known so far.
}

\keywords{
  Stars: individual: \object{WR\,118}, Stars: Wolf-Rayet, Stars: winds,
  outflows, Stars: circumstellar matter, Techniques: interferometric,
  Techniques: spectroscopic
}

\maketitle
%

\section{Introduction}\label{INTRO}

Since the work of \citet{ALLENETAL72}, infrared photometric studies of
Wolf-Rayet (WR) stars have shown that many late-type WC stars
have dust characteristics in addition to their typical free-free wind
emission.
For instance, \citet{WHT87} showed that 85\% of the WC9 stars and 50\%
of the WC8 stars are surrounded by heated ($T \sim 1300$~K)
circumstellar amorphous carbon dust \citep[see
also][]{CHIARETAL02
}. WR dust makers are rare but
remarkable in terms of their absolute dust-formation rate, which can be
as high as $\dot{M}=10^{-6}\,M_\odot{\rm yr}^{-1}$.

Among the strongest infrared WR sources, some show a distinct periodic
variability. Variable dust producers appear to be eccentric binary
systems (WR+OB) where episodic dust formation coincides with
periastron passage \citep[see, e.g.,][]{WH92}.
Although our knowledge of dusty WR stars has considerably increased in
the recent past,
we are still far from a complete understanding of how dust can form in
the hostile environment
of these hot stars.

Using aperture-masking interferometry with the Keck telescope, it has
been discovered that two supposedly single WRs with large IR excess,
\object{WR\,104} and \object{WR\,98a}, are close binary systems whose
dust distribution is tracing the orbital motion of the binary
components and forming a pattern which has been called ``pinwheel''
nebulae \citep{TUTHILLETAL99,MONNIERETAL99,MONNIERETAL07}.
%
%
Recently, the number of WR stars and, in particular, dusty WCs
discovered has considerably increased, especially in some open
clusters towards the galatic center
\citep{HUCHT06,CROWTHERETAL06,TUTHILLETAL06}. Among them, two new
pinwheel nebulae were discovered in the Quintuplet cluster 
\citep{TUTHILLETAL06}.


Binarity and wind clumping are currently the two best mechanisms to
form dust in the circumstellar environment of WR
stars. Observationally, distinguishing between these two cases is a
challenging task. Direct spectroscopic evidence for a binary companion
is difficult to reveal \citep{WH00}, but infrared long-baseline
interferometry with its spatial capability offers a unique chance to
resolve a central binary star and to determine the spatial
distribution of dust created in the circumstellar environment of
dust-enshrouded WR stars.

Recently, the dust shell of the persistent dust-maker \object{WR\,118}
(spectral type WC9), the third-brightest WR star in the $K$ band
($K=3.65^{\rm m}$), could be resolved with speckle
interferometry by \citet{YUDINETAL01} and \citet{MONNIERETAL07}. The
radiative transfer modeling of \citet{YUDINETAL01} showed that
grains can grow to sizes up to $\sim$0.6~$\mu$m.
They also found that the inner dust-shell rim diameter of \object{WR\,118}
is $r_{\rm in}=17\,$mas, 
temperature of the carbon-rich dust at the inner dust-shell
boundary is $1750\pm100\,$K, and dust formation rate is of the
order of $10^{-7}M_\odot{\rm yr}^{-1}$. Using the Keck telescope,
the aperture-masking observations of \citet{MONNIERETAL07} 
confirmed the results of \citet{YUDINETAL01} by measuring a Gaussian
FWHM of 23\,mas at 2.3$\mu$m.

Here, we present the first infrared long-baseline interferometric
observations of \object{WR\,118} obtained with AMBER, the near-infrared
beam-combiner instrument \citep{petrov07} of ESO's Very Large
Telescope  Interferometer (VLTI), giving access to spatial resolution
($\approx10$\,mas) five times better than before. This allowed us to
investigate the geometry of the innermost circumstellar dust shell of
\object{WR\,118}. 

\section{Observations and data processing}\label{OBS}

\object{WR\,118} was observed during three nights with AMBER in low
spectral resolution mode ($R=35$) in the $J$, $H$, and $K$ bands. The
telescope configuration used was E0-G0-H0 (linear array with baseline
lengths of 16/32/48\,m). Each measurement consists of 5000 frames with
an exposure time of 100~ms. We reduced the data with version 2.2
of the AMBER software package \texttt{amdlib}
\citep{2007A&A...464...29T}\footnote{provided by the Jean-Marie
  Mariotti Center, \url{http://www.jmmc.fr}}. Following previous AMBER
data reduction experience \citep[e.g., ][]{WEIGELT07}, we
kept 20\% of the frames with the highest fringe signal-to-noise ratio
and discarded frames with an optical-path difference offset larger
than 10$\,\mu$m. Due to limitations related to the weather
conditions and the extreme reddening of the object, the quality of the
$J$- and $H$-band data turned out to be too low for a scientific
analysis and were discarded. Thus, this paper discusses only the
$K$-band ($1.95-2.35\mu$m) observations of \object{WR\,118}. Also, we
could only extract visibilities from the shortest baseline out of the
2007 dataset. An overview of the AMBER observations is given in
Table~\ref{obslog}, and the corresponding UV coverage is shown in the
upper inset of Fig.~\ref{FigBestModelGeometric}.

\begin{table}[htbp]
  \vspace*{-3mm}
  \begin{tabular}{cccccc}
    \hline
    \multicolumn{6}{c}{object}\\
    \hline
    Date & Time & Name & $B_{\rm p}$        & PA      & seeing\\
    &    [UTC]      &            & [m]        & [$\degr$] & [$\arcsec$]   \\
    \hline
    2007-10-08    & 02:18:53  & \object{WR\,118}& 9.2    & 74.3      & 0.61\\
    2008-03-28    & 08:43:43  & \object{WR\,118}&13.4-40.2    & 61.7      & 0.83\\
    2008-03-31    & 06:51:05  & \object{WR\,118}& 9.6-28.8    & 42.6      & 0.73\\
    2008-03-31    & 07:09:33  & \object{WR\,118}&10.3-32.0    & 47.6      & 0.86\\
    \hline
    \multicolumn{6}{c}{ calibrators}\\
    \hline
    Date& Time &  HD        &  UD        &  spectral &  seeing\\
    &    [UTC]      &  number    & [mas]     & type     & [\arcsec]   \\
    \hline
    2007-10-08    & 01:07:58  & \object{175583}   &  $1.02\pm 0.01$   & K2III    & 0.66\\
    2008-03-28    & 09:33:52  &\object{ 175583}   &  $1.02\pm 0.01$   & K2III    & 0.60 \\
    2008-03-31    & 07:42:46  & \object{175583}   &  $1.02\pm 0.01$   & K2III    & 0.81 \\
    2008-03-31    & 08:20:04  & \object{165524}   &  $1.11\pm 0.01$   & K3III    & 0.80 \\
    \hline
  \end{tabular}
  \caption[]{\label{obslog}
    Log of the AMBER observations of \object{WR\,118} (top) and
    the corresponding calibrators (bottom). 
    $B_{\rm p}$ gives the range of projected baselines, and PA denotes
    the position angle of the observation. The calibrators'
    uniform-disk diameters were taken from \citet{charm2}
    and the spectral types are from Simbad
    (\url{http://simbad.u-strasbg.fr}). 
  }
  \vspace*{-5mm}
\end{table}


The calibrated AMBER visibilities and closure phases of
\object{WR\,118} are shown in Fig.~\ref{FigBestModelGeometric} as a
function of spatial frequency $q$. The figure reveals a visibility
decrease up to $q=60$, followed by a ``bump'' around $q=80$. This
visibility shape is characteristic for an object intensity
distribution that contains sharp edges. Examples of such intensity
distributions are, e.g., a uniform disk or a ring-like intensity
distribution. In addition, the AMBER data clearly show non-zero,
non-180$^\circ$ closure phases, indicating that the circumstellar
envelope deviates from a point-symmetric configuration. Therefore,
from this qualitative analysis of the AMBER data, we can clearly rule
out any centro-symmetric shape of the object; i.e., the circumstellar
shell cannot be described by a simple circular or even elliptical
envelope, as was done, for instance, in \citet{YUDINETAL01}.

\section{Modeling}\label{MODEL}

\subsection{Simple geometrical models}\label{GEOMOD}

To interpret the AMBER data, we first tried to fit them with
simple two-component geometrical models: one
resolved component (Gaussian disk, uniform disk, or ring) to account
for the overall visibility decrease towards larger spatial
frequencies, and one spatially off-set component (resolved or
not) to account for asymmetry, mainly probed  by the non-zero closure
phase. We assumed that the object's intensity distribution is not
  wavelength-dependent in order to take advantage of the larger UV
coverage. We performed a global optimization of the fit by
using a simulated annealing algorithm, complemented with standard
gradient descent \citep[see ][for a first use of these algorithms]{Millour09}.

\begin{figure}[htbp]
  \vspace*{-4mm}
  \centering
  \includegraphics[width=0.40\textwidth]{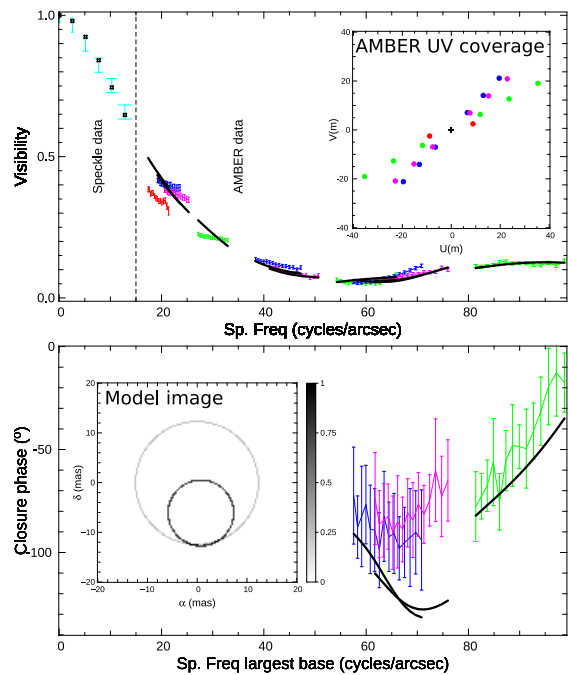}
  \caption{
    AMBER visibilities (top, points with error bars, UV
    coverage in the inset) and closure phases (bottom) of WR\,118,
    compared to the best-fitting model. The
    \citet{YUDINETAL01} data (light blue) are shown
    in the top panel. The best-fitting geometrical model made of two
    rings and an extended background component (40\,mas FWHM, not
    sketched) is shown as black lines (model image in the inset of
    the bottom panel).
  }
  \label{FigBestModelGeometric}
  \vspace*{-7mm}
\end{figure}

\begin{table}[htbp]
  \vspace*{-2mm}
  \begin{center}
    \begin{tabular}{ccc}
      \hline
      Parameter  & Description &  Value \\
      \hline
      S1  &  1$^{st}$ ring size & $24.8\pm3.1$\,mas \\
      F2  &  2$^{nd}$ ring flux fraction & $34\pm7$\%  \\
      R. A.   & 2$^{nd}$ ring R. A.-offset & $0.79\pm1.7$\,mas  \\
      Dec.   & 2$^{nd}$ ring Dec.-offset & $-6.0\pm2.3$\,mas  \\
      S2  & 2$^{nd}$ ring size & $13.2\pm1.0$\,mas  \\
      F3  & Background flux fraction  & $46\pm11$\%  \\
      \hline
    \end{tabular}
    \caption{\label{table:parametersSimpleModel}
      Parameters of our best-fitting geometrical model composed of two
      rings and an extended background component. 
    }
  \end{center}
  \vspace*{-7mm}
\end{table}


The reduced $\chi^2$ of the solution was typically 50 depending on the
models we tried to fit to the AMBER data. Therefore, all these models
are probably an inappropriate representation of the measured
visibilities and phases.
Thus, in the next modeling step, we tentatively added a new, fully
resolved component that dilutes the flux of the whole system.
 By including this additional fully resolved component, the reduced
 $\chi^2$ for each model was considerably decreased (down to about 20).
The results of our fitting procedure can be summarized as follows:



It is not possible to find a reasonable fit to the data
with only a two-component model. A significant improvement of the fit
is achieved by adding a fully resolved background component
to the two-component models, accounting for $\approx50$\% of the total
flux. Thus, \object{WR\,118}, as seen by AMBER, is composed of at least a fully
resolved structure plus a component with a more complex shape, at
smaller scales.

All the models have a typical size of $\approx$20\,mas, which is
approximately in agreement with the results of \citet{YUDINETAL01} and
\citet{MONNIERETAL07}.

The models containing an extended component plus a point
source locate the point source inside the extension of the resolved
component. If this point source was a star and the resolved component
a dust shell, then the putative companion star would be located inside
the dust shell. Then, the companion star would have
probably carved out the dust shell. Therefore, the hypothesis of
having a continuous dust shell plus a well-separated companion can
be discarded.
%

 Our best-fitting model, which is able to reproduce both the
  visibilities and closure phases, consists of a set of two rings, with
  the smaller one located inside the larger one and with a contact
  zone in the South (see Table~\ref{table:parametersSimpleModel} and
  Fig.~\ref{FigBestModelGeometric}). Best-fitting models using more
  than two rings also show interweaved rings of  increasing sizes.

Next, we provide a physical interpretation to describe the WR118
AMBER data, taking into account the knowledge acquired from the
simple geometrical models.

\begin{figure}[htbp]
  \vspace*{-5mm}
  \centering
  \includegraphics[height=0.42\textwidth, angle=-90]{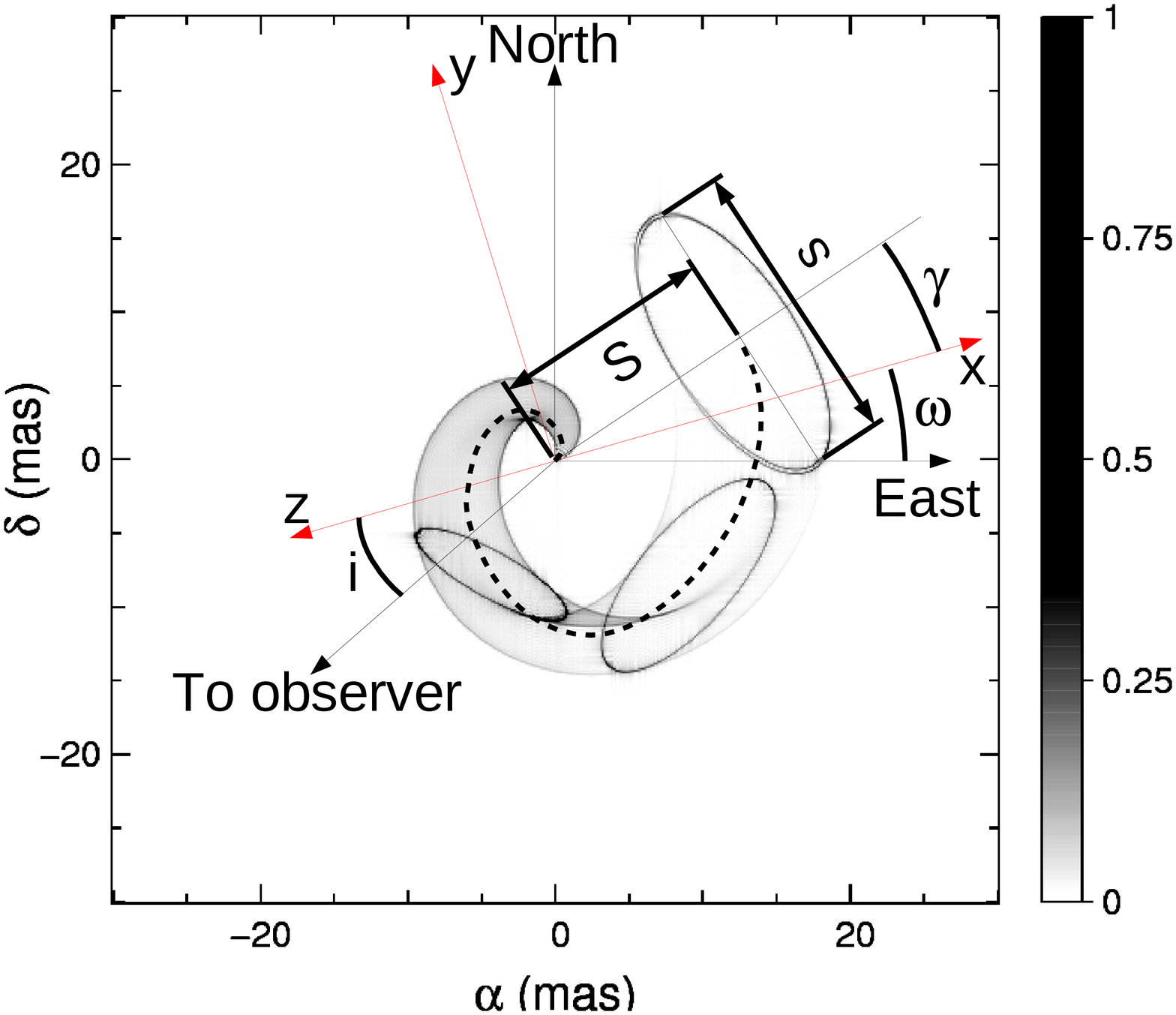}\\
\vspace*{-2mm}	
  \includegraphics[height=0.45\textwidth, angle=-90]{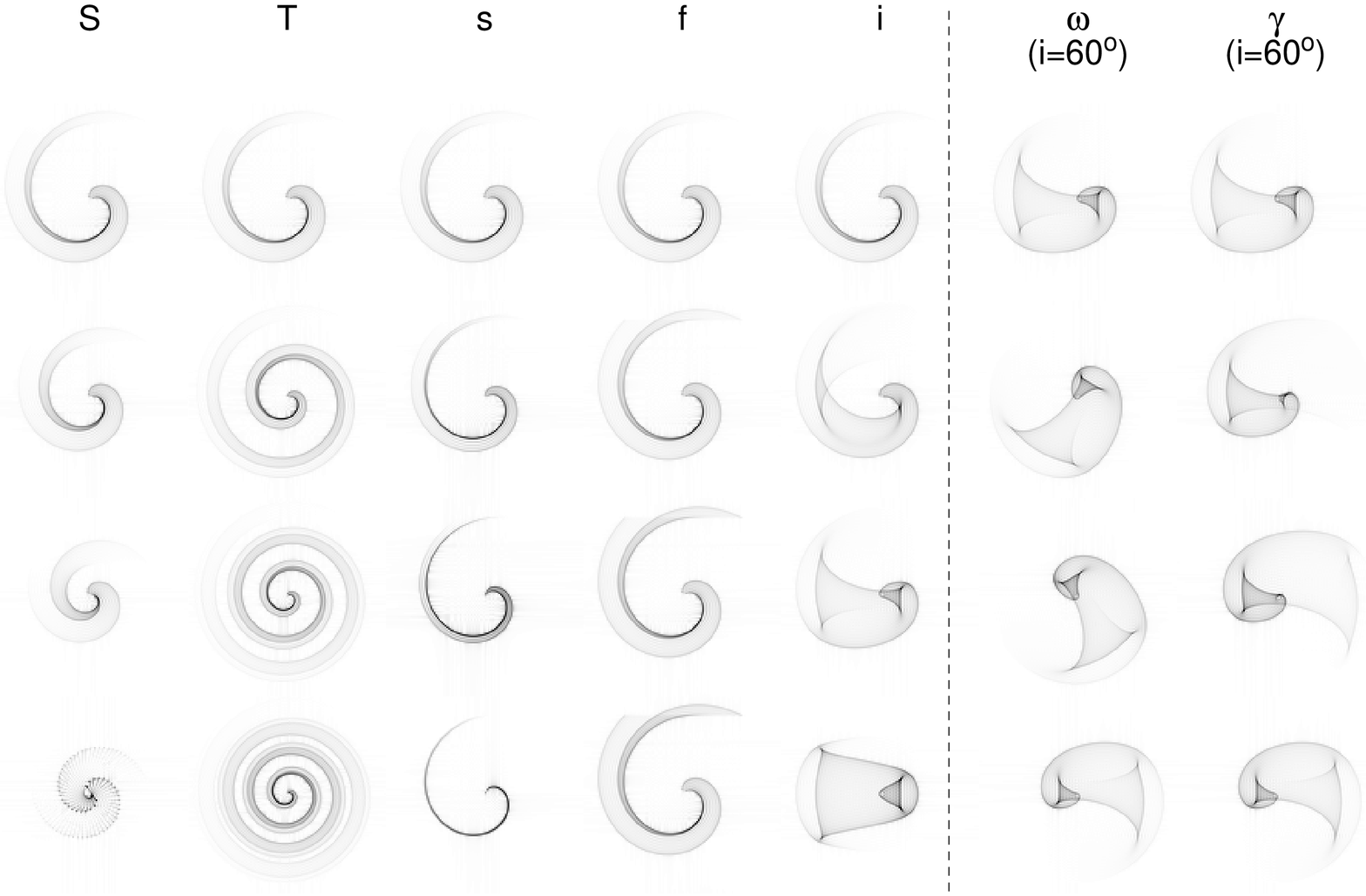}
\vspace*{-2mm}	
  \caption{
    Top: 
    Sketch of our pinwheel nebula model. A series of elliptical rings
    (three of them are highlighted) following a projected
    Archimedian spiral (dashed line) set it up. The parameters of the
    model are labeled in the sketch.
    Bottom:
    Dependence of the pinwheel model on its parameters. For the first
    five parameters, the same model is shown in the first row. For the
    two last, a model with an inclination of $60^\circ$ is shown.
  }
  \label{FigPinWheelModel}
  \vspace*{-2mm}
\end{figure}

\subsection{Pinwheel model}

\citet{TUTHILLETAL99} and \citet{MONNIERETAL99,MONNIERETAL07} showed
that the dusty WR stars \object{WR\,98a} and \object{WR\,104} are
surrounded by spiral-like dusty nebulae. In particular, they found
that the dust plume around these stars follow an Archimedian spiral,
projected onto the plane of sky. \citet{HARRIES04} presented a
radiative transfer model of a bow-shock cavity between the winds of
the two stellar components and successfully compared it with the
observations.

The goal of this section is to demonstrate that a pinwheel
nebula is consistent with all the features seen in the AMBER
data of \object{WR\,118}. To explain these features, we developed a
geometrical model of a pinwheel nebula, based on the same simple
geometrical models presented in the previous section. It is generated
by adding many shifted elliptical rings of increasing sizes
representing the bow-shock cavities at different orbital phases of
the binary star along an Archimedian spiral, and then subsequently
projecting them onto the plane of sky, as illustrated in the top panel of
Fig.~\ref{FigPinWheelModel}. Such a pinwheel model can be fully
described by the following eight parameters:

\begin{itemize}
\item the extension $S$, and the number of turns $T$ of the spiral,
\item the size of the last ring $s$, corresponding to the opening
  angle $\theta = \arctan [s/S]$ of the wind-wind collision zone,
\item the flux fraction $f$ of the last ring relative to the central one; 
  the ring flux in the spiral linearly decreases from 1 to $f$,
\item the three projection angles: $i$, $\omega$ and $\gamma$, to
  project the pinwheel on the plane of sky,
\item $F$, the fractional flux of a fully resolved background.
\end{itemize}

To illustrate the meaning of the individual input parameters, in the
bottom panel of Fig.~\ref{FigPinWheelModel}, we show how the morphology
of the pinwheel changes if a specific parameter is changed.

Following the results of \citet{TUTHILL08}, who showed that for
\object{WR\,104}, the pinwheel flux sharply drops after exactly one
turn around the binary star, we decided to fix $T=1$ and $f=0$ (see
Table \ref{PINWHEEL}). Therefore, for our pinwheel model, we
use the same number of free parameters (i.e. six) as in
Sect.~\ref{GEOMOD}.

\begin{figure}[htbp]
	\vspace*{-3mm}	
  \centering
  \includegraphics[width=0.40\textwidth]{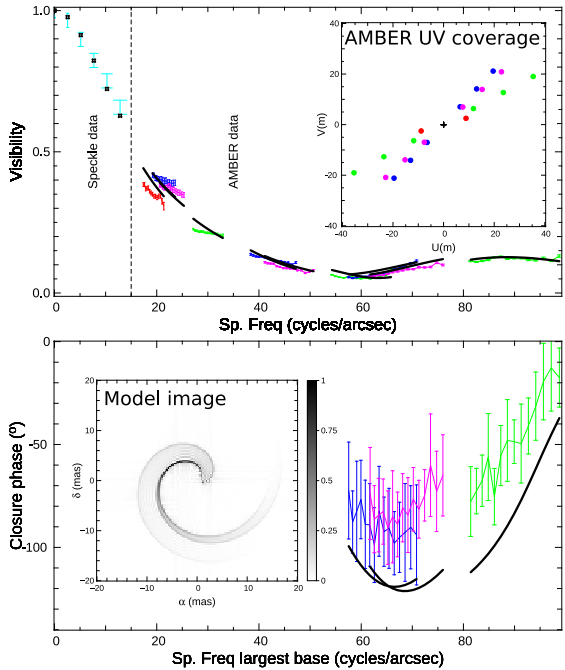}
  \caption{
    Same as Fig.~\ref{FigBestModelGeometric}, but showing
    the best-fitting pinwheel model (black solid lines). An image of
    our best-fitting pinwheel model is shown in the bottom panel's
    inset.
  }
  \label{FigVisibPinwheel}
   \vspace*{-0mm}
\end{figure}

The best-fitting pinwheel model parameters are summarized in
Table~\ref{PINWHEEL}, and the comparison with the AMBER data is shown
in Fig.~\ref{FigVisibPinwheel}. The qualitative
agreement of the fit is very similar to the two-ring model
described in the previous section, and the reduced $\chi^2$
is also similar ($\chi^2=16$). 
Nevertheless, such a pinwheel model is much easier to understand from
a physical point-of-view, for instance, in terms of a dust-emission
signature in the wind-wind curved shock around a binary star.

Some parameters are well-constrained by our fit, such as the total
extension of the pinwheel $S=13.9\pm1.1$\,mas, the extension of
the extreme ring $s=22.6\pm4.7$\,mas, or the background flux
contribution $F=49\pm5$\%. On the other hand, other parameters
like the projection angles onto the plane of sky and especially the
inclination angle are poorely constrained. This may come from
the tight range of position angles of the AMBER data.

\begin{table}[htbp]
   \vspace*{-2mm}
  \begin{center}
    \begin{tabular}{ccc}
      \hline
      Parameter  & Description &  Value \\
      \hline
      S  & Total size & $13.9\pm1.1$\,mas \\
      T  & Turns & 1 (fixed) \\
      s  & Ring-size at extremity & $22.6\pm4.7$\,mas  \\
      f  & Flux at extremity & 0 (fixed)  \\
      i  & Inclination angle & $3\pm26$\deg  \\
      $\omega$  & Orientation on sky & $9.2\pm9.6$\deg  \\
      $\gamma$  & Rotation angle & $0.0\pm9.6$\deg  \\
      F  & Background flux fraction & $49\pm5$\% \\
      \hline
    \end{tabular}
    \caption{\label{PINWHEEL}
      Parameter of our best-fitting pinwheel model (see also
      Fig.~\ref{FigVisibPinwheel}).
    }
  \end{center}
  \vspace*{-7mm}
\end{table} 

We also found good agreement between our models and both the
AMBER and \citet{YUDINETAL01} data sets by using a size for the
extended component of $\approx40$\,mas (see
Figs.~\ref{FigBestModelGeometric} and \ref{FigVisibPinwheel}), which is
fully resolved by AMBER. 

\subsection{Spherical wind clumping model}

A second physical hypothesis to form dust in WR stars is wind clumping
in single star winds \citep{2000A&A...357..572C}. We simulated a
spherical clumped envelope by randomly putting unresolved ``clumps''
along a Gaussian profile with $\approx30$\,mas FWHM. 
 If we assume a small number of clumps ($\leq10$), the
  model visibilities hardly match the observed ones, while some clump
  configurations match the observed closure phases.
 For many clumps \citep[$\approx1000$, hypothesis favored by,
   e.g.,][]{2000AJ....120.3201L}, the model visibilities more easily
   match the AMBER ones, while, on average, the model closure phases
   show values closer to zero (or 180$^\circ$) than the observations.

Thus, from this study we qualitatively conclude that spherical
wind-clumping models difficultly account for both the AMBER
visibilities and closure phases.

\section{Discussion and conclusion}

Our study showed that our \object{WR\,118} data can be described by a
dusty pinwheel nebula model. Other types of models describe the data
equally well, but spherical wind-clumping models cannot easily explain
both the observed visibilities and closure phases. Therefore, we
suggest that WR118 is a binary star embedded in a pinwheel-like dust
nebula.


Assuming the pinwheel hypothesis, we can measure the opening angle of the
bow-shock of WR118: $\theta = \arctan [s/S] = 58\pm7\degr$. This is
comparable to $40\degr$ in the case of WR104 \citep{TUTHILL08}. We
also infer a period for the putative binary system in WR\,118,
assuming, like in WR\,98a \citep{MONNIERETAL99}, tidal circularization
of the orbit. Given a typical wind terminal velocity of a WC9 star of
1200\,kms$^ {-1}$ \citep{2007ARA&A..45..177C}, a distance of
$\approx$3\,kpc for WR\,118 \citep{2001NewAR..45..135V}, and an
extension of 13.9\,mas, the free-flying time of a dust plume from the
center of the system to its farthest elongation is $\approx$60\,days,
corresponding to one full rotation of the binary in our model.

Such a period is remarkably shorter than the periods
of 241.5\,days for WR\,104 \citep{TUTHILL08}, 565\,days for WR\,98a
\citep{MONNIERETAL99} and 24.8\,yrs for WR\,112
\citep{MARCHENKOETAL02}. This would make WR\,118 the shortest-period
pinwheel system known today.

The short period found for WR\,118 may also be compared to other known
WR binaries. For instance, $\gamma^2$ Velorum is a WC8+O
star with a similar period, of 78\,days \citep{MILLOUR07}, but it does
not produce dust. WR118 has a later spectral type than $\gamma^2$
Velorum, which alters the chemistry of the wind. Alternatively, the
distance of WR\,118 might be severely underestimated or its terminal
velocity might be slower than for a typical WC9 star. 

With the study presented in this paper, we demonstrated the potential
of infrared long-baseline interferometry to resolve pinwheel nebulae
around dusty WR stars. In \citet{2001NewAR..45..135V, HUCHT06}, we can
find 24 permanent dust-producing WR stars, of which five (not yet
including WR\,118) have been confirmed to be colliding-wind binaries
through the direct detection of pinwheel nebulae. Eight have
near-infrared sizes strongly in favor of pinwheel nebulae
\citep{TUTHILLETAL06, MONNIERETAL07} and are just waiting to be
confirmed. In total, there are 18 dusty WR stars still to be resolved
by long-baseline interferometry in order to provide observational
evidence that all WR permanent dust-maker are colliding winds binaries.

\begin{acknowledgements}
  
  We thank the ESO VLTI team on Paranal and in Garching for carrying
  out the AMBER observations presented in this paper. The data
  were reduced using the publicly available data
  reduction software package {\it amdlib}, kindly provided by the
  Jean-Marie Mariotti Center (\url{http://www.jmmc.fr}).
  %

\end{acknowledgements}
\bibliographystyle{aa}
\bibliography{WR118}

\begin{thebibliography}{25}
\expandafter\ifx\csname natexlab\endcsname\relax\def\natexlab#1{#1}\fi

\bibitem[{{Allen} {et~al.}(1972){Allen}, {Swings}, \& {Harvey}}]{ALLENETAL72}
{Allen}, D.~A., {Swings}, J.~P., \& {Harvey}, P.~M. 1972, \aap, 20, 333

\bibitem[{{Cherchneff} {et~al.}(2000){Cherchneff}, {Le Teuff}, {Williams}, \&
  {Tielens}}]{2000A&A...357..572C}
{Cherchneff}, I., {Le Teuff}, Y., {Williams}, P., \& {Tielens}, A. 2000, \aap,
  357, 572

\bibitem[{{Chiar} {et~al.}(2002){Chiar}, {Peeters}, \& {Tielens}}]{CHIARETAL02}
{Chiar}, J.~E., {Peeters}, E., \& {Tielens}, A. 2002, \apjl, 579, L91

\bibitem[{{Crowther}(2007)}]{2007ARA&A..45..177C}
{Crowther}, P. 2007, \araa, 45, 177

\bibitem[{{Crowther} {et~al.}(2006){Crowther}, {Hadfield}, {Clark},
  {Negueruela}, \& {Vacca}}]{CROWTHERETAL06}
{Crowther}, P., {Hadfield}, L., {Clark}, J., {Negueruela}, I., \& {Vacca}, W.
  2006, \mnras, 372, 1407

\bibitem[{{Harries} {et~al.}(2004){Harries}, {Monnier}, {Symington}, \&
  {Kurosawa}}]{HARRIES04}
{Harries}, T., {Monnier}, J., {Symington}, N., \& {Kurosawa}, R. 2004, \mnras,
  350, 565

\bibitem[{{L{\'e}pine} {et~al.}(2000){L{\'e}pine}, {Moffat}, {St-Louis},
  {Marchenko}, {Dalton}, {Crowther}, {Smith}, {Willis}, {Antokhin}, \&
  {Tovmassian}}]{2000AJ....120.3201L}
{L{\'e}pine}, S., {Moffat}, A.~F.~J., {St-Louis}, N., {et~al.} 2000, \aj, 120,
  3201

\bibitem[{{Marchenko} {et~al.}(2002){Marchenko}, {Moffat}, {Vacca},
  {C{\^o}t{\'e}}, \& {Doyon}}]{MARCHENKOETAL02}
{Marchenko}, S., {Moffat}, A., {Vacca}, W., {C{\^o}t{\'e}}, S., \& {Doyon}, R.
  2002, \apjl, 565, L59

\bibitem[{{Millour} {et~al.}(2009){Millour}, {Chesneau}, {Borges Fernandes},
  Meilland, Mars, Benoist, Thi\'ebaut, Stee, Hofmann, Baron, Young, Bendjoya,
  Carciofi, de~Souza, Driebe, Jankov, Kervella, Petrov, Robbe-Dubois, Vakili,
  Waters, \& Weigelt}]{Millour09}
{Millour}, F., {Chesneau}, O., {Borges Fernandes}, M., {et~al.} 2009, A\&A,
  accepted.

\bibitem[{{Millour} {et~al.}(2007){Millour}, {Petrov}, {Chesneau}, {Bonneau},
  {Dessart}, {Bechet}, {Tallon-Bosc}, {Tallon}, {Thi{\'e}baut}, {Vakili},
  {Malbet}, {Mourard}, {Antonelli}, {Beckmann}, {Bresson}, {Chelli},
  {Dugu{\'e}}, {Duvert}, {Gennari}, {Gl{\"u}ck}, {Kern}, {Lagarde}, {Le
  Coarer}, {Lisi}, {Perraut}, {Puget}, {Rantakyr{\"o}}, {Robbe-Dubois},
  {Roussel}, {Tatulli}, {Weigelt}, {Zins}, {Accardo}, {Acke}, {Agabi},
  {Altariba}, {Arezki}, {Aristidi}, {Baffa}, {Behrend}, {Bl{\"o}cker},
  {Bonhomme}, {Busoni}, {Cassaing}, {Clausse}, {Colin}, {Connot},
  {Delboulb{\'e}}, {Domiciano de Souza}, {Driebe}, {Feautrier}, {Ferruzzi},
  {Forveille}, {Fossat}, {Foy}, {Fraix-Burnet}, {Gallardo}, {Giani}, {Gil},
  {Glentzlin}, {Heiden}, {Heininger}, {Hernandez Utrera}, {Hofmann}, {Kamm},
  {Kiekebusch}, {Kraus}, {Le Contel}, {Le Contel}, {Lesourd}, {Lopez}, {Lopez},
  {Magnard}, {Marconi}, {Mars}, {Martinot-Lagarde}, {Mathias}, {M{\`e}ge},
  {Monin}, {Mouillet}, {Nussbaum}, {Ohnaka}, {Pacheco}, {Perrier}, {Rabbia},
  {Rebattu}, {Reynaud}, {Richichi}, {Robini}, {Sacchettini}, {Schertl},
  {Sch{\"o}ller}, {Solscheid}, {Spang}, {Stee}, {Stefanini}, {Tasso}, {Testi},
  {von der L{\"u}he}, {Valtier}, {Vannier}, \& {Ventura}}]{MILLOUR07}
{Millour}, F., {Petrov}, R.~G., {Chesneau}, O., {et~al.} 2007, \aap, 464, 107

\bibitem[{{Monnier} {et~al.}(1999){Monnier}, {Tuthill}, \&
  {Danchi}}]{MONNIERETAL99}
{Monnier}, J., {Tuthill}, P., \& {Danchi}, W. 1999, \apjl, 525, L97

\bibitem[{{Monnier} {et~al.}(2007){Monnier}, {Tuthill}, {Danchi}, {Murphy}, \&
  {Harries}}]{MONNIERETAL07}
{Monnier}, J., {Tuthill}, P., {Danchi}, W., {Murphy}, N., \& {Harries}, T.
  2007, \apj, 655, 1033

\bibitem[{{Petrov} {et~al.}(2007){Petrov}, {Malbet}, {Weigelt}, {Antonelli},
  {Beckmann}, {Bresson}, {Chelli}, {Dugu{\'e}}, {Duvert}, {Gennari},
  {Gl{\"u}ck}, {Kern}, {Lagarde}, {Le Coarer}, {Lisi}, {Millour}, {Perraut},
  {Puget}, {Rantakyr{\"o}}, {Robbe-Dubois}, {Roussel}, {Salinari}, {Tatulli},
  {Zins}, {Accardo}, {Acke}, {Agabi}, {Altariba}, {Arezki}, {Aristidi},
  {Baffa}, {Behrend}, {Bl{\"o}cker}, {Bonhomme}, {Busoni}, {Cassaing},
  {Clausse}, {Colin}, {Connot}, {Delboulb{\'e}}, {Domiciano de Souza},
  {Driebe}, {Feautrier}, {Ferruzzi}, {Forveille}, {Fossat}, {Foy},
  {Fraix-Burnet}, {Gallardo}, {Giani}, {Gil}, {Glentzlin}, {Heiden},
  {Heininger}, {Hernandez Utrera}, {Hofmann}, {Kamm}, {Kiekebusch}, {Kraus},
  {Le Contel}, {Le Contel}, {Lesourd}, {Lopez}, {Lopez}, {Magnard}, {Marconi},
  {Mars}, {Martinot-Lagarde}, {Mathias}, {M{\`e}ge}, {Monin}, {Mouillet},
  {Mourard}, {Nussbaum}, {Ohnaka}, {Pacheco}, {Perrier}, {Rabbia}, {Rebattu},
  {Reynaud}, {Richichi}, {Robini}, {Sacchettini}, {Schertl}, {Sch{\"o}ller},
  {Solscheid}, {Spang}, {Stee}, {Stefanini}, {Tallon}, {Tallon-Bosc}, {Tasso},
  {Testi}, {Vakili}, {von der L{\"u}he}, {Valtier}, {Vannier}, \&
  {Ventura}}]{petrov07}
{Petrov}, R.~G., {Malbet}, F., {Weigelt}, G., {et~al.} 2007, \aap, 464, 1

\bibitem[{{Richichi} {et~al.}(2005){Richichi}, {Percheron}, \&
  {Khristoforova}}]{charm2}
{Richichi}, A., {Percheron}, I., \& {Khristoforova}, M. 2005, \aap, 431, 773

\bibitem[{{Tatulli} {et~al.}(2007){Tatulli}, {Millour}, {Chelli}, {Duvert},
  {Acke}, {Hernandez Utrera}, {Hofmann}, {Kraus}, {Malbet}, {M{\`e}ge},
  {Petrov}, {Vannier}, {Zins}, {Antonelli}, {Beckmann}, {Bresson}, {Dugu{\'e}},
  {Gennari}, {Gl{\"u}ck}, {Kern}, {Lagarde}, {Le Coarer}, {Lisi}, {Perraut},
  {Puget}, {Rantakyr{\"o}}, {Robbe-Dubois}, {Roussel}, {Weigelt}, {Accardo},
  {Agabi}, {Altariba}, {Arezki}, {Aristidi}, {Baffa}, {Behrend}, {Bl{\"o}cker},
  {Bonhomme}, {Busoni}, {Cassaing}, {Clausse}, {Colin}, {Connot},
  {Delboulb{\'e}}, {Domiciano de Souza}, {Driebe}, {Feautrier}, {Ferruzzi},
  {Forveille}, {Fossat}, {Foy}, {Fraix-Burnet}, {Gallardo}, {Giani}, {Gil},
  {Glentzlin}, {Heiden}, {Heininger}, {Kamm}, {Kiekebusch}, {Le Contel}, {Le
  Contel}, {Lesourd}, {Lopez}, {Lopez}, {Magnard}, {Marconi}, {Mars},
  {Martinot-Lagarde}, {Mathias}, {Monin}, {Mouillet}, {Mourard}, {Nussbaum},
  {Ohnaka}, {Pacheco}, {Perrier}, {Rabbia}, {Rebattu}, {Reynaud}, {Richichi},
  {Robini}, {Sacchettini}, {Schertl}, {Sch{\"o}ller}, {Solscheid}, {Spang},
  {Stee}, {Stefanini}, {Tallon}, {Tallon-Bosc}, {Tasso}, {Testi}, {Vakili},
  {von der L{\"u}he}, {Valtier}, \& {Ventura}}]{2007A&A...464...29T}
{Tatulli}, E., {Millour}, F., {Chelli}, A., {et~al.} 2007, \aap, 464, 29

\bibitem[{{Tuthill} {et~al.}(1999){Tuthill}, {Monnier}, \&
  {Danchi}}]{TUTHILLETAL99}
{Tuthill}, P., {Monnier}, J., \& {Danchi}, W. 1999, \nat, 398, 487

\bibitem[{{Tuthill} {et~al.}(2006){Tuthill}, {Monnier}, {Tanner}, {Figer},
  {Ghez}, \& {Danchi}}]{TUTHILLETAL06}
{Tuthill}, P., {Monnier}, J., {Tanner}, A., {et~al.} 2006, Science, 313, 935

\bibitem[{{Tuthill} {et~al.}(2008){Tuthill}, {Monnier}, {Lawrance}, {Danchi},
  {Owocki}, \& {Gayley}}]{TUTHILL08}
{Tuthill}, P., {Monnier}, J.~D., {Lawrance}, N., {et~al.} 2008, \apj, 675, 698

\bibitem[{{van der Hucht}(2001)}]{2001NewAR..45..135V}
{van der Hucht}, K.~A. 2001, New Astronomy Review, 45, 135

\bibitem[{{van der Hucht}(2006)}]{HUCHT06}
{van der Hucht}, K.~A. 2006, \aap, 458, 453

\bibitem[{{Weigelt} {et~al.}(2007){Weigelt}, {Kraus}, {Driebe}, {Petrov},
  {Hofmann}, {Millour}, {Chesneau}, {Schertl}, {Malbet}, {Hillier}, {Gull},
  {Davidson}, {Domiciano de Souza}, {Antonelli}, {Beckmann}, {Bresson},
  {Chelli}, {Dugu{\'e}}, {Duvert}, {Gennari}, {Gl{\"u}ck}, {Kern}, {Lagarde},
  {Le Coarer}, {Lisi}, {Perraut}, {Puget}, {Rantakyr{\"o}}, {Robbe-Dubois},
  {Roussel}, {Tatulli}, {Zins}, {Accardo}, {Acke}, {Agabi}, {Altariba},
  {Arezki}, {Aristidi}, {Baffa}, {Behrend}, {Bl{\"o}cker}, {Bonhomme},
  {Busoni}, {Cassaing}, {Clausse}, {Colin}, {Connot}, {Delboulb{\'e}},
  {Feautrier}, {Ferruzzi}, {Forveille}, {Fossat}, {Foy}, {Fraix-Burnet},
  {Gallardo}, {Giani}, {Gil}, {Glentzlin}, {Heiden}, {Heininger}, {Hernandez
  Utrera}, {Kamm}, {Kiekebusch}, {Le Contel}, {Le Contel}, {Lesourd}, {Lopez},
  {Lopez}, {Magnard}, {Marconi}, {Mars}, {Martinot-Lagarde}, {Mathias},
  {M{\`e}ge}, {Monin}, {Mouillet}, {Mourard}, {Nussbaum}, {Ohnaka}, {Pacheco},
  {Perrier}, {Rabbia}, {Rebattu}, {Reynaud}, {Richichi}, {Robini},
  {Sacchettini}, {Sch{\"o}ller}, {Solscheid}, {Spang}, {Stee}, {Stefanini},
  {Tallon}, {Tallon-Bosc}, {Tasso}, {Testi}, {Vakili}, {von der L{\"u}he},
  {Valtier}, {Vannier}, {Ventura}, {Weis}, \& {Wittkowski}}]{WEIGELT07}
{Weigelt}, G., {Kraus}, S., {Driebe}, T., {et~al.} 2007, \aap, 464, 87

\bibitem[{{Williams} \& {van der Hucht}(1992)}]{WH92}
{Williams}, M. \& {van der Hucht}, K.~A. 1992, in A. S. P. Conf. Series,
  Vol.~22, 269

\bibitem[{{Williams} \& {van der Hucht}(2000)}]{WH00}
{Williams}, P.~M. \& {van der Hucht}, K.~A. 2000, \mnras, 314, 23

\bibitem[{{Williams} {et~al.}(1987){Williams}, {van der Hucht}, \&
  {The}}]{WHT87}
{Williams}, P.~M., {van der Hucht}, K.~A., \& {The}, P.~S. 1987, \aap, 182, 91

\bibitem[{{Yudin} {et~al.}(2001){Yudin}, {Balega}, {Bl{\"o}cker}, {Hofmann},
  {Schertl}, \& {Weigelt}}]{YUDINETAL01}
{Yudin}, B., {Balega}, Y., {Bl{\"o}cker}, T., {et~al.} 2001, \aap, 379, 229

\end{thebibliography}

\end{document}